\documentclass[preprint]{aastex}

\def\url#1{{\ttfamily\def\/{/\discretionary{}{}{}}#1}}

\def\mathnew{\mathsurround=0pt}
\def\simov#1#2{\lower .5pt\vbox{\baselineskip0pt
    \lineskip-.5pt\ialign{$\mathnew#1\hfil##\hfil$\crcr#2\crcr\sim\crcr}}}  
\def\simgreat{\mathrel{\mathpalette\simov >}}

\def\'#1{\ifx#1i{\accent"13\i}\else{\accent"13#1}\fi}

\shorttitle{DM and DE vs.~CMB data}
\shortauthors{Mainini et al.}
\begin{document}
\title{Cosmologies with dynamical and coupled Dark Energy vs.
CMB data}

\author{Silvio A. Bonometto, Luciano Casarini, Loris P.L. Colombo \&
Roberto Mainini }

\affil{Physics Department G. Occhialini, Universit\`a degli Studi di
Milano--Bicocca, Piazza della Scienza 3, I20126 Milano (Italy) 
\& I.N.F.N., Sezione di Milano}

\begin{abstract}
We compare a large set of cosmologies with WMAP data, performing a fit
based on a MCMC algorithm.  Besides of $\Lambda$CDM models, we take
dynamical DE models, where DE and DM are uncoupled or coupled, both in
the case of constant coupling and in the case when coupling varies
with suitable laws. DE however arises from a scalar field
self--interacting through a SUGRA potential.  We find that the best
fitting model is SUGRA dynamical DE, almost indipendently from the
exponent $\alpha$ in the self--interaction potential.

The main target of this work are however coupled DE models, for which
we find limits on the DE--DM coupling strength. In the case of
variable coupling, we also find that greater values of the Hubble
constant are preferred.

\end{abstract}

\keywords{cosmology: theory -- dark energy }


\section{Introduction}
Cosmologies with $\Omega_{o,de} \simeq 0.75$, $\Omega_{o,m} \simeq
0.25$, $\Omega_{o,b} \simeq 0.03$, $h \simeq 0.7$ and $n_s\simeq1$ fit
available data on CMB, LSS, and SNIa. Respectively, the above symbols
are: the {\it present} density parameters for dark energy (DE), whole
non--relativistic matter, baryons, the Hubble parameter in units of
100 km/s/Mpc, and the primeval spectral index. CMB is the Cosmic
Microwave Background, LSS is the Large Scale Structure. Possible
references for the above data are: Tegmark et al. (2001), De Bernardis
et al. (2000), Hanany et al. (2000), Halverson et al. (2002), Spergel
et al. (2003), Percival et al. (2002), Efstathiou et al. (2002), Riess
et al. (1998), Perlmutter et al.  (1999).

The nature of DE is not clear. This paper is devoted to comparing
various hypotheses on its nature among them and with WMAP data. It is
already known that, in order to fit data, DE must have a largely
negative pressure: $p_{de} = w\rho_{de}$, with $-w \sim 0.7$--1
($\rho_{de}$ is DE energy density). False vacua obey such a state
equation with $w=-1$, but assuming DE to arise from false vacuum
implies a {\it fine--tuning} of $\rho_{de}$, at the end of the EW
phase transition by $1:10^{\, 56}$. Otherwise, DE could be due to a
self--interacting scalar field $\phi$ (Wetterich 1988, Ratra \&
Peebles 1988, RP hereafter). In this case we have {\it dynamical} DE
(dDE) or quintessence.  

The principal analysis of WMAP first--year data \citep{Spergel}
constrained flat $\Lambda$CDM models defined by six parameters:
$\Omega_{o,b} h^2$, $\Omega_{o,m} h^2$, $h$, $n_s$, the fluctuation
amplitude $A$ and the optical depth $\tau$.  As possible extensions of
$\Lambda$CDM cosmologies, several works considered models with a fixed
state parameter $w \equiv p_{de} / \rho_{de}$
(Spergel et al. 2003, Bean et al. 2004, Tegmark et al. 2004, Melchiorri et al. 2004),
or adopted
$z$--dependent parameterizations of $w(z)$ interpolating between
early--time and late--time values
(Corasaniti et al. 2004, Jassal et al. 2005, Rapetti et al. 2004).
A general conclusion was
that current data mostly allow to constrain only the present state
parameter, $w(z=0) \lesssim -0.80$. Here we shall extend the
comparison to dDE models admitting a tracker solution.

Possible interactions of DE were also considered in the literature.
Severe limits can be set to its interaction with baryons; Amendola
(2000, 2003), Gasperini, Piazza \& Veneziano (2002), Perrotta \&
Baccigalupi (2002), among others, discussed models where it interacts
with Dark Matter (DM).  Amendola \& Quercellini (2003)  performed a
preliminary study on the limits on DE--DM interactions, arising from
WMAP data, for DE theories with constant coupling (ccDE).  It is also
possible to consider cosmologies where the DM--DE coupling constant
$C$, that we shall define below in accordance with Amendola (2000), is
replaced by $1/\phi$ and is therefore variable in time (vcDE). A
physical origin for these models was recently discussed by Mainini \&
Bonometto (2003; see also Colombo et al. 2003).

In this paper we perform a systematic fit of WMAP data to these
different cosmologies: dDE, ccDE, vcDE, and $\Lambda$CDM. For the
first 3 of them a SUGRA self--interaction potential will be taken
(Brax \& Martin 1999, 2000; Brax, Martin \& Riazuelo 2000); its
expression is given in Sec.~2$\, $, together with essential
information on coupled and uncoupled DE models. Further information on
coupled models can be found, {\it e.g.}, in Colombo et al. (2004).

The fit is performed by using a MCMC (MonteCarlo Markov Chain)
algorithm, in the same way as the WMAP team fit the $\Lambda$CDM
model. In Sec.~3, we describe the fitting procedure.
In the case of that model we re--obtain their results.

The number of parameters to be fitted depends on the nature of the
model.  $\Lambda$CDM and vcDE models have the same number of
parameters.  In dDE models there is one extra parameter and ccDE
models have a further parameter in respect to the previous cathegory.
In evaluating the success of a fit, however, the number of parameters
is taken into account. In Sec.~4 we perform sich comparison and in
Sec.~5 we draw some conclusions.

\section{The SUGRA potential}
\begin{figure}
\vskip -1.truecm
\plotone{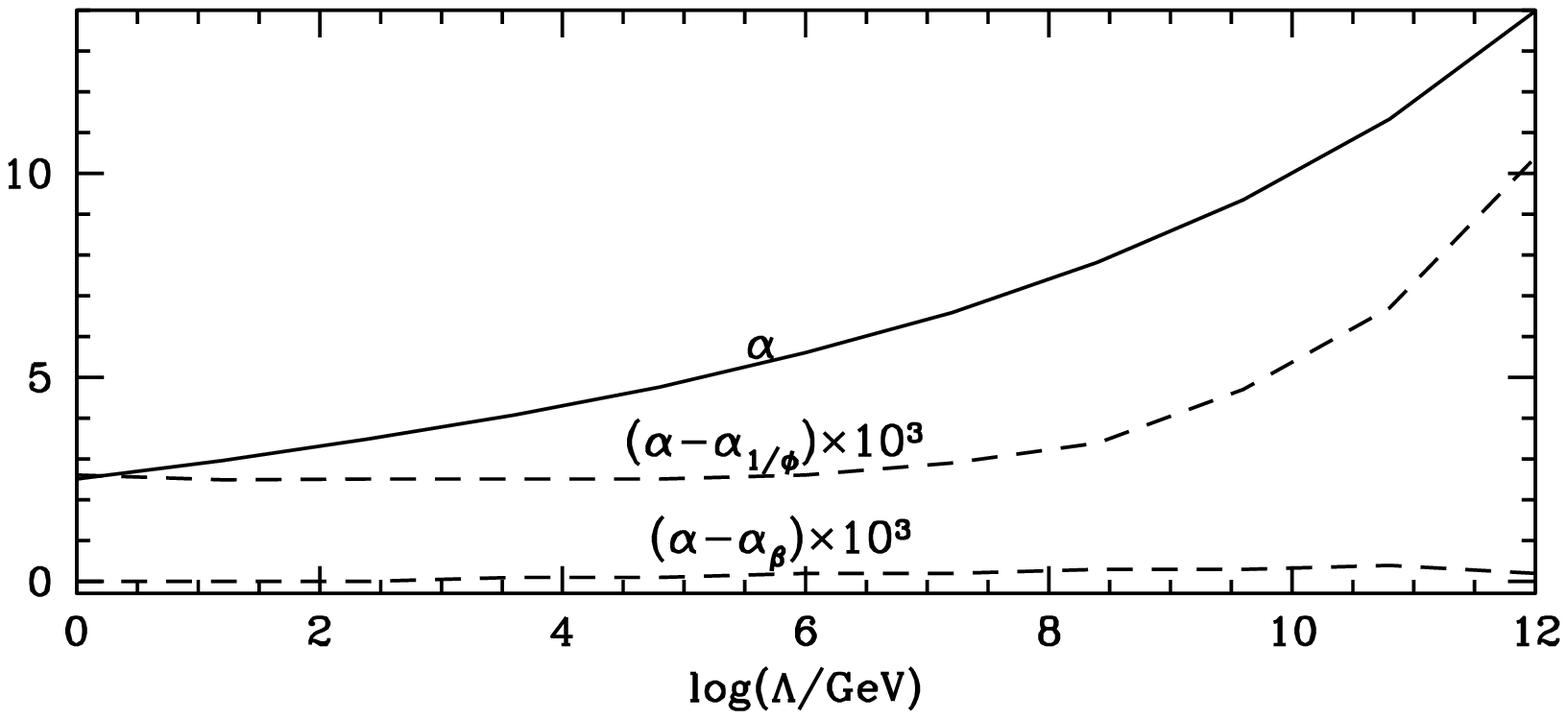}
\epsscale{1.0}
\vskip -9.7truecm
\caption{Values of $\alpha$ corresponding to given $\Lambda$ scales in
SUGRA models with $\Omega_{o,dm} = 0.27$. The tiny differences, just
above noise level, for coupled models ($\alpha_\beta$ and
$\alpha_{1/\phi}$ for constant coupling and $\phi^{-1}$ coupling,
respectively), are also shown.}
\vskip .1truecm
\label{fal}
\end{figure}

All models with dynamical DE we consider make use of a SUGRA potential
\begin{equation}
V(\phi) = {\Lambda^{\alpha+4} \over \phi^\alpha} 
\exp \left( 4 \pi\phi^2 \over m_p^2 \right)
\label{eq:l1}
\end{equation}
to yield the self--interaction of the scalar field $\phi$; here $m_p =
G^{-1/2}$ is the Planck mass; $V$ depends on the two parameters
$\Lambda$ and $\alpha$. When they are assigned, $\Omega_{de}$ is
however fixed. We therefore prefere to use $\Omega_{de}$ and $\Lambda$
as independent parameters. The latter scale is related to $\alpha$ as
shown in Figure \ref{fal}, which holds also in the presence of
DM--DE coupling.
The potential appears in the Lagrangian
\begin{equation}
{\cal L} = \sqrt{-g} \left\{ {1 \over 2} 
g_{\mu\nu} \partial_\mu \phi \partial_\nu \phi 
- V(\phi) \right\} ~;
\label{m1}
\end{equation}
in the absence of coupling, the equation of motion
\begin{equation}
\ddot \phi + 2 {\dot a \over a}  \dot \phi + 
a^2  V'(\phi) = 0
\label{eq:m4}
\end{equation}
follows, while the expressions for energy density $\rho_{\phi} =
\rho_{\phi;kin} + \rho_{\phi; pot}$ and pressure $p_{\phi} =
\rho_{\phi;kin} - \rho_{ \phi;pot}$ are obtainable by combining the terms
\begin{equation}
\rho_{\phi,kin} = {\dot \phi^2 \over 2 a^2} ~,~~
\rho_{\phi,pot} = V(\phi)~.
\label{enpre}
\end{equation}
In the radiation dominated era, eq.~(\ref{eq:m4}) admits the
tracker solution
\begin{equation}
\phi^{\alpha+2} = g_\alpha \Lambda^{\alpha+4} a^2 \tau^2 ~,
\label{eq:l2}
\end{equation}
with $g_\alpha = \alpha (\alpha+2)^2/4(\alpha+6)$. This tracker
solution fairly defines the initial conditions also in the presence of
coupling, although, in the presence of coupling, the equation of
motion becomes
\begin{equation}
\ddot \phi + 2 {\dot a \over a}  \dot \phi + 
a^2  V'(\phi) = C a^2 \rho_c 
\label{couple}
\end{equation}
while CDM energy density varies according to the equation
\begin{equation}
\dot \rho_c + 3 {\dot a \over a} \rho_v = 
-  C a^2 \rho_c ~;
\end{equation}
baryon and radiation equations keep the usual shape.
The coefficient $C$ sets the coupling strength and, 
in the case of constant coupling (ccDE), we can set
\begin{equation}
C = \left(\pi \over 3 \right)^{1 \over 2} {4 \over m_p } \beta
\end{equation}
so to parametrize the coupling strength through the adimensional
coefficient $\beta$. Here we consider also a $\phi$--dependent
coupling (vcDE)
\begin{equation}
C = \phi^{-1}
\end{equation}
for which there is no extra parameter to fix the coupling strength.

\section{Comparison with WMAP data through a MCMC algorithm}
WMAP data have been extensively used to constrain cosmological
parameters. They are high precision estimates of the anisotropy power
spectrum $C_l^T$ up to $l \sim 900$, as well of the TE correlation
power spectrum $C_l^{TE}$ up to $l \sim 450$. We shall fit these data
to possible cosmologies, by considering a parameter space of 6 to 8
dimensions. A grid-based likelihood analysis would require a huge CPU
time and we use a Markov Chain Monte Carlo (MCMC) approach, as it has
become customary for CMB analysis 
(Christensen et al. 2001, Knox et al. 2001, Lewis et al. 2002, Kosowski et al. 2002, Dunkley et al. 2004).

To this aim, as in any attempt to fit CMB data to models, a linear
code providing $C_l$'s is needed. Here we use our optimized extension
of CMBFAST \citep{cmbfast}, able to inspect also ccDE and vcDE
cosmologies. The likelihood of each model is then evaluated through
the publicly available code by the WMAP team \citep{wmap:verde} and
accompanying data \citep{wmap:hinshaw,wmap:kogut}.

A MCMC algorithm samples a known distribution ${\cal L}({\bf x})$ by
means of an arbitrary trial distribution $p({\bf x})$. Here $\cal L$
is a likelihood and ${\bf x}$ is a point in the parameter space. The
chain is started from a random position ${\bf x}$ and moves to a new
position ${\bf x}^\prime$, according to the trial distribution. The
probability of accepting the new point is given by ${\cal L}({\bf
x}^\prime)/{\cal L}({\bf x})$; if the new point is accepted, it is
added to the chain and used as the starting position for a new
step. If ${\bf x}^\prime$ is rejected, a replica of ${\bf x}$ is added
to the chain and a new ${\bf x}^\prime$ is tested.

In the limit of infinitely long chains, the distribution of points
sampled by a MCMC describes the underlying statistical process. Real
chains, however, are finite and convergence criteria are critical.
Moreover, a chain must be required to fully explore the high
probability region in the parameter space. Statistical properties
estimated using a chain which has yet to achieve good convergence or
mixing may be misleading. Several methods exist to diagnose mixing and
convergence, involving either single long chains or multiple chains
starting from well separated points in the parameter space, as the one
used here. Once a chain passes convergence tests, it is an accurate
representation of the underlying distribution.

In order to ensure mixing, we run six chains of $\sim 30000$ points
each, for each model category. We diagnose convergence by requiring
that, for each parameter, the variances both of the single chains and
of the whole set of chains ($W$ and $B$, respectively) satisfy the
Gelman \& Rubin test~\cite{wmap:verde,gelrub}, $R < 1.1$ with:
\begin{equation}
R = {[(N-1)/N]W +(1 +1/M)B \over W}~. 
\end{equation}
Here each chain has $2N$ points, but only the last $N$ points are used
to estimate variances, and $M$ is the total number of chains. In most
model categories considered, we find that the slowest parameter to
converge is $\lambda$.
 
\section{Results}
\begin{figure}
\plotone{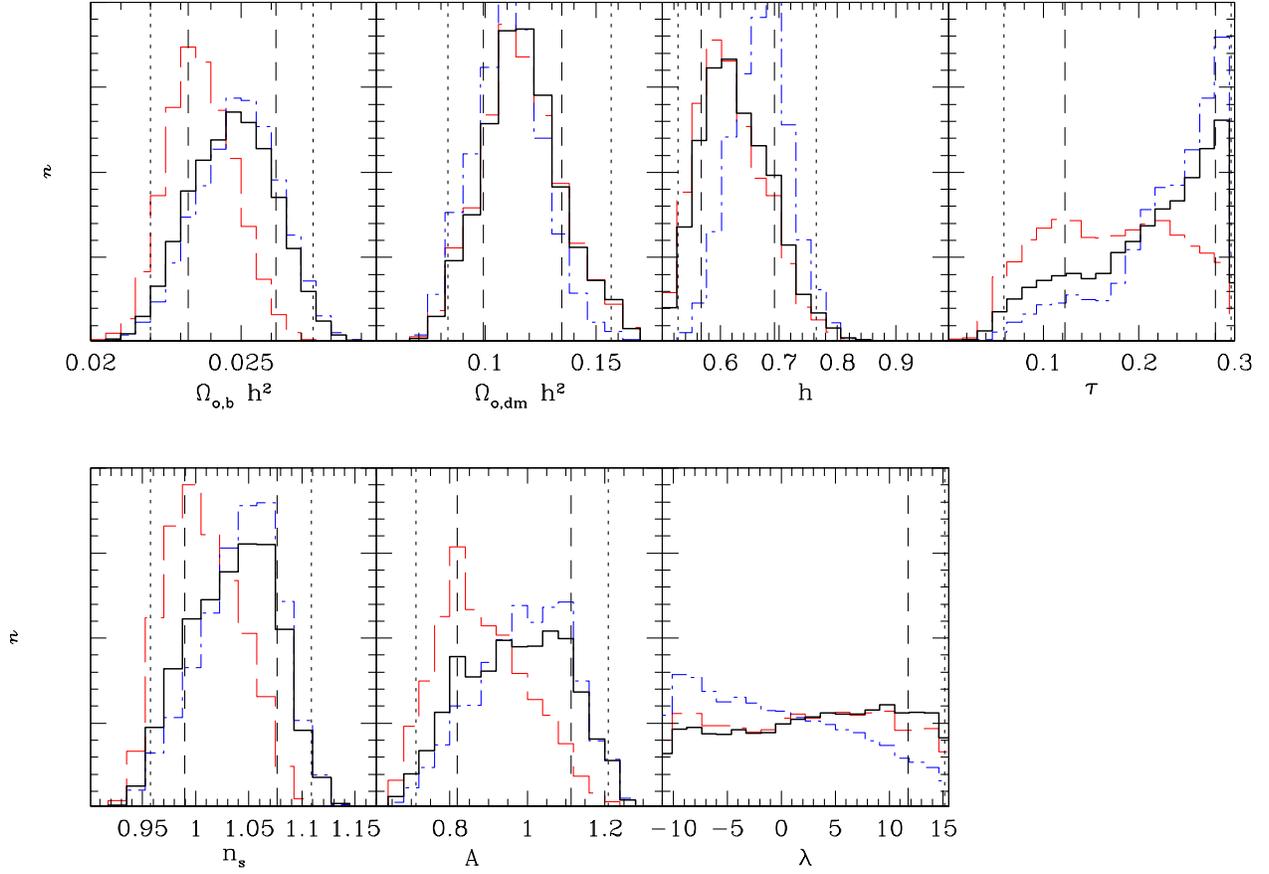}
\epsscale{1.0}
\caption{Marginalized distributions for the 7--parameters SUGRA model
with no priors (solid lines), BBNS prior (long dashed) or HST prior
(dot--dashed).  Short dashed (dotted) vertical lines show the
boundaries of 68.3 \% c.l.  (95.4 \% c.l.) interval; for $\lambda$
only upper limits are shown.}
\label{fig:pdf1D1}
\end{figure}

\begin{figure}
\plotone{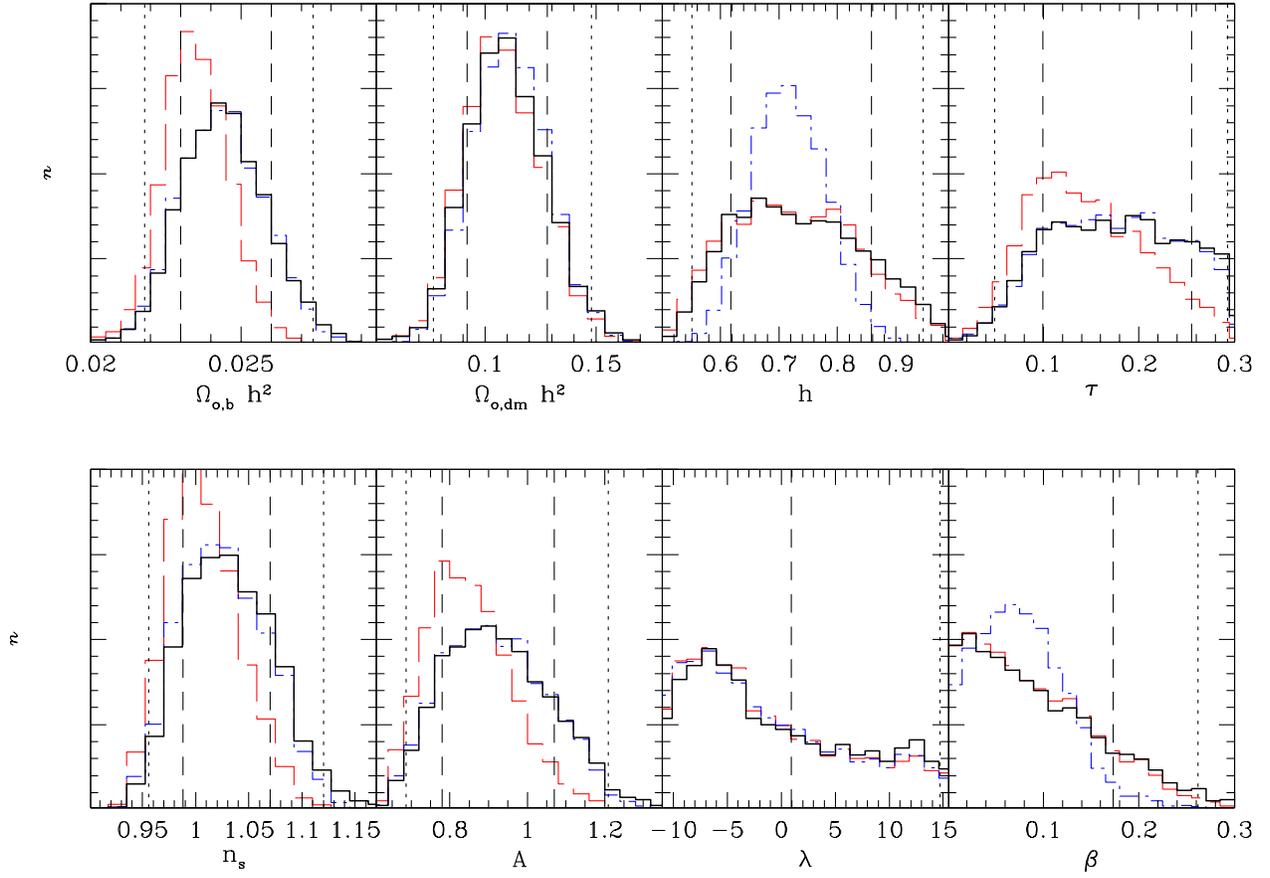}
\epsscale{1.0}
\caption{As Fig.~\ref{fig:pdf1D1} but for the 8--parameters ccDE
model. For $\lambda$ and $\beta$ only the upper c.l. boundaries are
shown.}
\label{fig:pdf1D2}
\end{figure}

\begin{figure}
\plotone{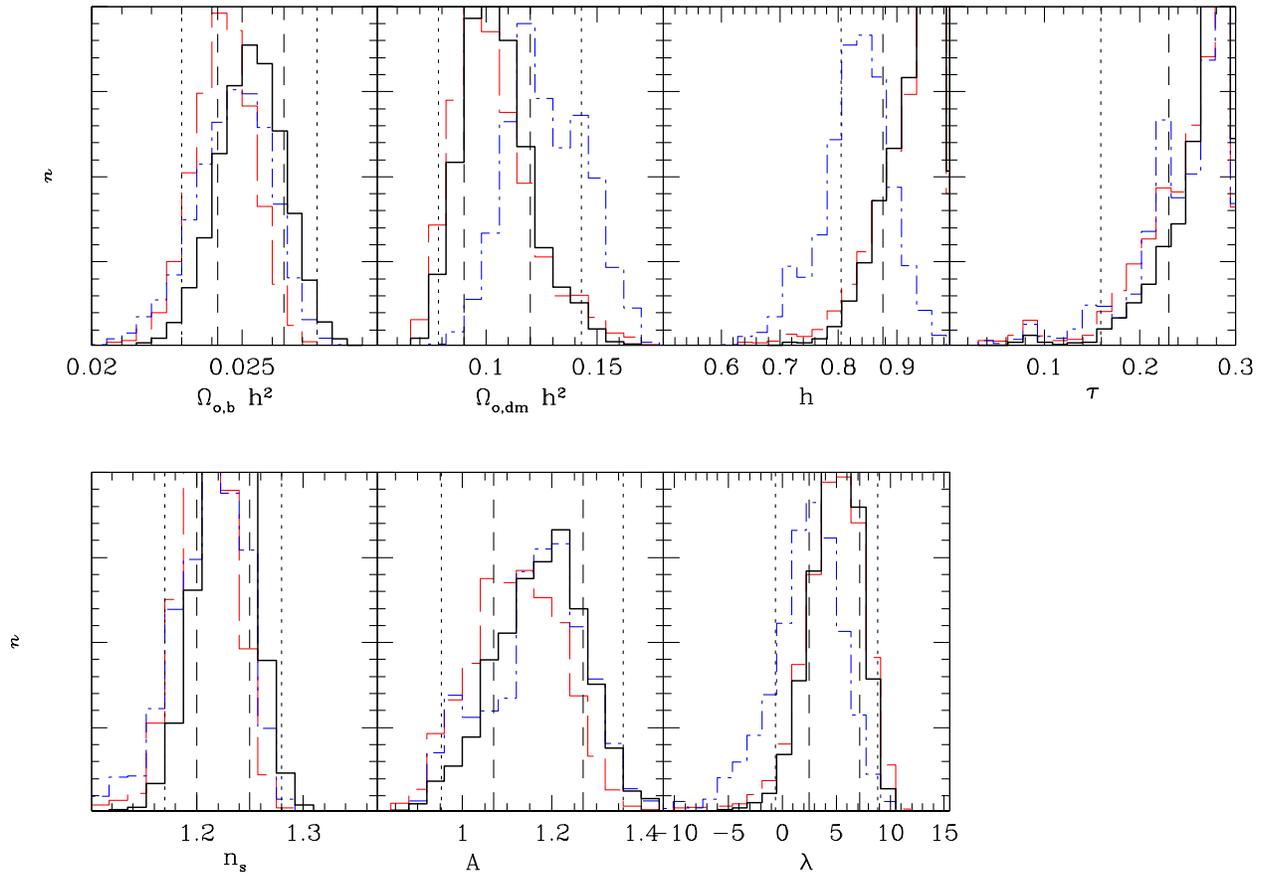}
\epsscale{1.0}
\caption{As Fig.~\ref{fig:pdf1D1} but for the 7--parameters vcDE
model.}
\label{fig:pdf1D3}
\end{figure}

\begin{figure}
\plotone{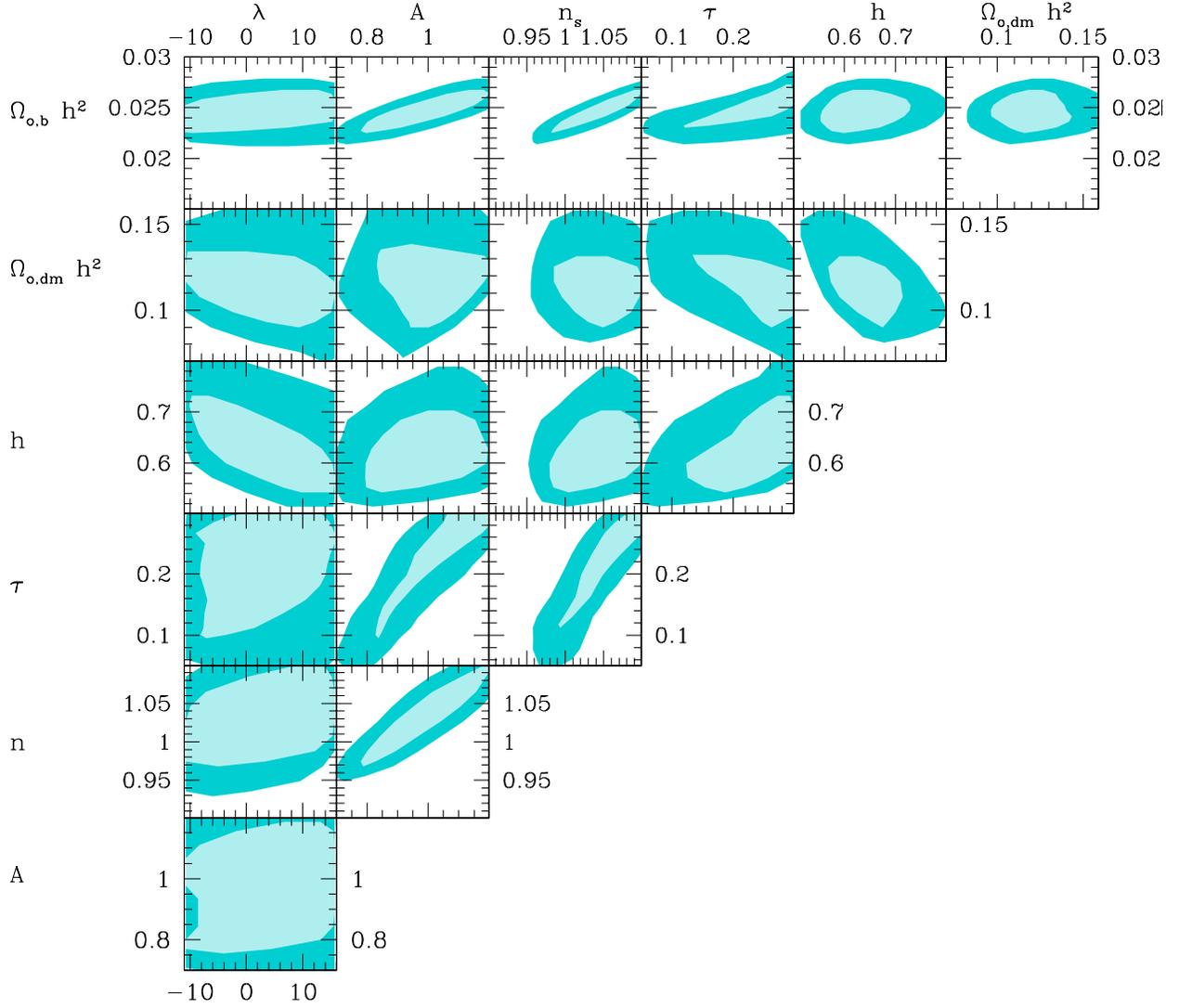}
\epsscale{1.0}
\caption{Joint 2D constraints for SUGRA models. 
Light (dark) shaded areas delimit the region enclosing 68.3 \%
(95.4 \%) of the total points.}
\label{fig:pdf2D1}
\end{figure}

\begin{figure}
\plotone{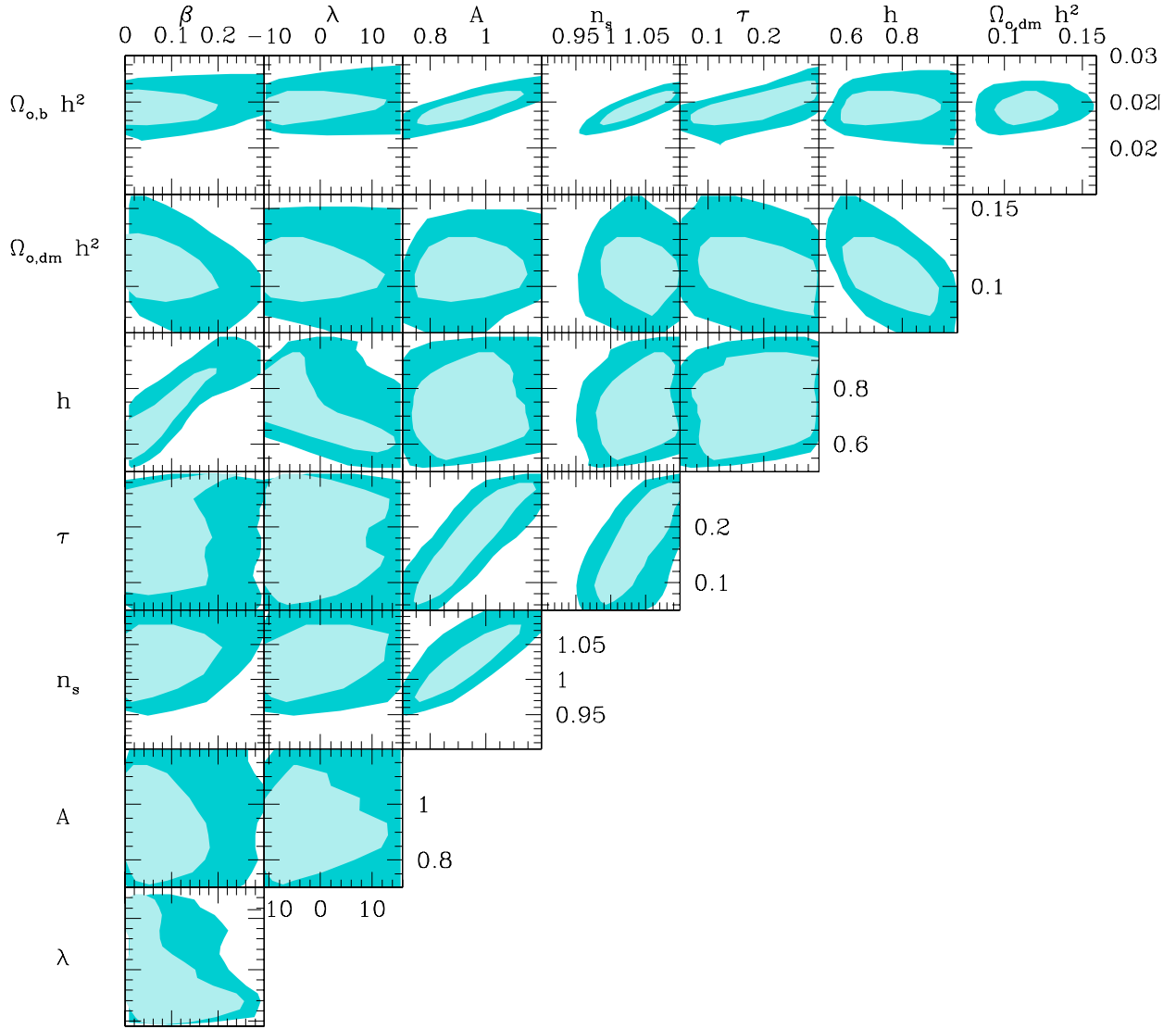}
\epsscale{1.0}
\caption{As Fig.~\ref{fig:pdf2D1} but for ccDE models.}
\label{fig:pdf2D2}
\end{figure}

\begin{figure}
\plotone{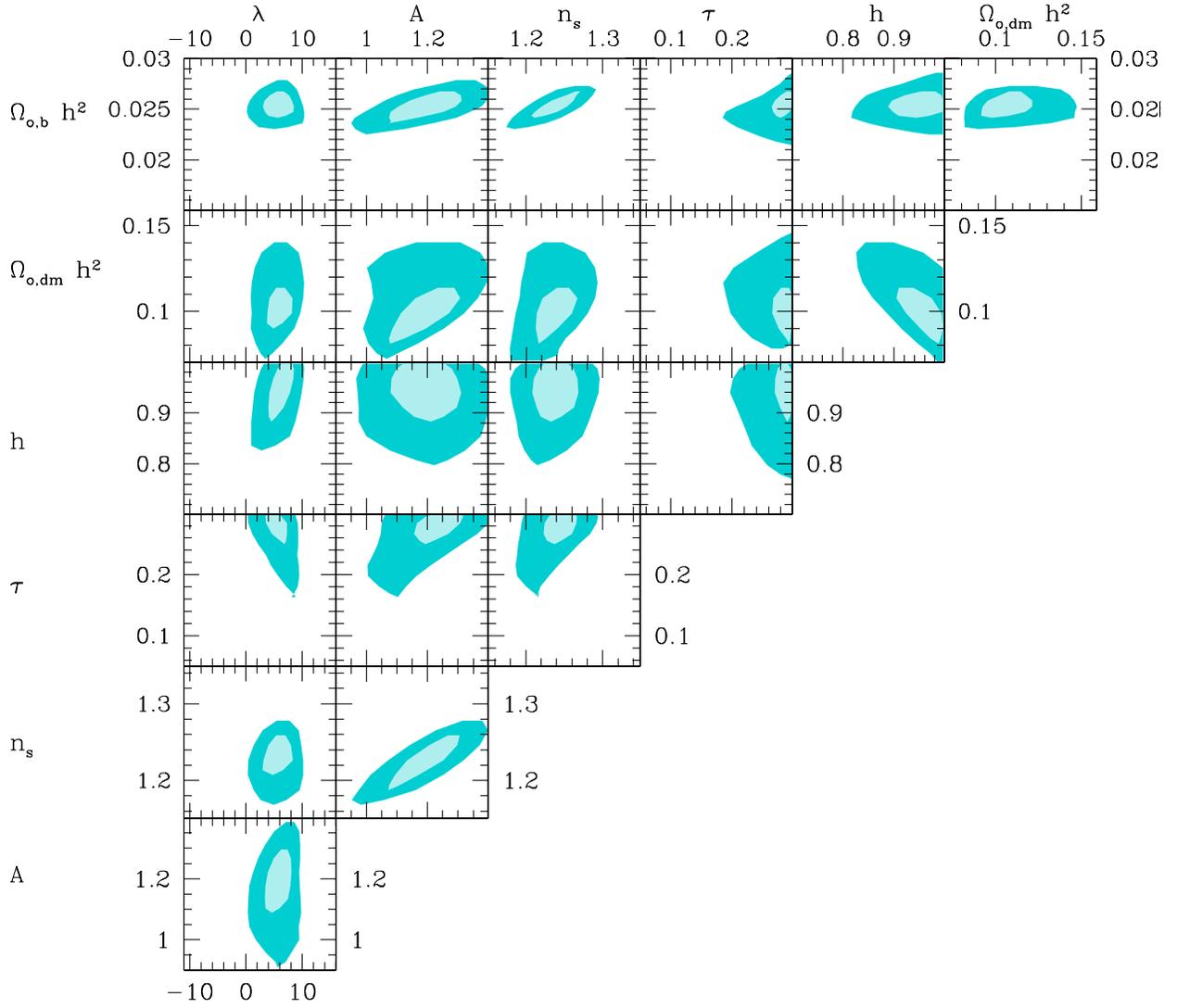}
\epsscale{1.0}
\caption{As Fig.~\ref{fig:pdf2D1} but for vcDE models. Here,
cosmological parameters are more stringently constrained than in other
models.}
\label{fig:pdf2D3}
\end{figure}

The basic results of this work are shown in the Tables
\ref{tab:res1}--\ref{tab:res3}. For each model category we provide the
expectation values and the variance of each parameter, as well as
the values of the parameters of the best fitting models.  The
corresponding marginalized distributions are plotted in
figures~\ref{fig:pdf1D1}--\ref{fig:pdf1D3}, while joint 2D confidence
regions are shown in figures~\ref{fig:pdf2D1}--\ref{fig:pdf2D3}.

The values of $\chi^2$ can also be compared, taking into account the
number of degrees of freedom. This is shown in Table~\ref{tab:chi2}.
The smallest $\chi^2$ is obtained for the uncoupled SUGRA model, which
performs slightly better than $\Lambda$CDM.

Tables 1 and 2 and the corresponding figures, {\it concerning
uncoupled or constant--coupling} SUGRA models, show that WMAP data
scarcely constrain $\lambda=\log(\Lambda/ {\rm GeV})$, allowed to vary
from $\sim -12$ to 16. On the contrary, in the case of a $\phi^{-1}$
coupling, loose but precise limitations on $\lambda$ are set, as is
shown in Table 3. In the presence of this coupling, the 2--$\sigma$
$\Lambda$--interval ranges from $\sim 10$ to $\sim 3 \cdot
10^{10}$GeV.
\begin{table}
\caption{SUGRA parameters for dDE models: for each parameter $x$, the
expectation value $\langle x \rangle$, its variance $\sigma_x$, and
its maximum likelihood value $x_{max}$, in the 7--dimensional
parameter space, are shown.}
\label{tab:res1}
\vglue 0.2truecm
\begin{center}
\begin{tabular}{cccc}
\hline
\rule[-1ex]{0pt}{3.5ex}
      $ x $    &  $ \langle x \rangle$  &  $\sigma_x$ &  $x_{max}$  
  \\
\hline
$\Omega_{o,b} h^2$  &     0.025  &   0.001  &    0.026  \\
$\Omega_{o,dm} h^2$  &     0.12   &   0.02   &    0.11   \\
$ h $           &     0.63   &   0.06   &    0.58   \\
$ \tau$         &     0.21   &   0.07   &    0.28   \\
$ n_s$          &     1.04   &   0.04   &    1.08   \\
$ A $           &     0.97   &   0.13   &    1.11   \\
$\lambda$       &     3.0    &    7.7   &    13.7   \\
\hline
\end{tabular}
\end{center}
\vglue -0.5truecm
\caption{SUGRA parameters for ccDE models; the parameter space is
7--dimensional and parameter are shown as in the previous Table.}
\label{tab:res2}
\vglue0.2truecm
\begin{center}
\begin{tabular}{cccc}
\hline
\rule[-1ex]{0pt}{3.5ex}
      $ x $    &  $ \langle x \rangle$  &  $\sigma_x$ &  $x_{max}$  
  \\
\hline
$\Omega_{o,b} h^2$  &     0.024  &   0.001  &    0.024  \\
$\Omega_{o,dm} h^2$  &     0.11   &   0.02   &    0.12   \\
$ h $           &     0.74   &   0.11   &    0.57   \\
$ \tau$         &     0.18   &   0.07   &    0.17   \\
$ n_s$          &     1.03   &   0.04   &    1.02   \\
$ A $           &     0.92   &   0.14   &    0.93   \\
$\lambda$       &    -0.5    &    7.6   &    8.3    \\
$\beta$         &     0.10   &   0.07   &    0.07   \\
\hline
\end{tabular}
\end{center}
\vglue -0.5truecm
\caption{SUGRA parameters for vcDE models. At variance from other
model categories, $\lambda$ here is constrained.  Parameter values are
shown as in Table 1.}
\label{tab:res3}
\vglue0.2truecm
\begin{center}
\begin{tabular}{cccc}
\hline
\rule[-1ex]{0pt}{3.5ex}
      $ x $    &  $ \langle x \rangle$  &  $\sigma_x$ &  $x_{max}$  
  \\
\hline
$\Omega_{o,b} h^2$  &     0.025  &   0.001  &    0.026  \\
$\Omega_{o,dm} h^2$  &     0.11   &   0.02   &    0.09   \\
$ h $           &     0.93   &   0.05   &    0.98   \\
$ \tau$         &     0.26   &   0.04   &    0.29   \\
$ n_s$          &     1.23   &   0.04   &    1.23   \\
$ A $           &     1.17   &   0.10   &    1.20   \\
$\lambda$       &     4.8    &    2.4   &     5.7   \\
\hline
\end{tabular}
\end{center}
\end{table}
\begin{table}
\caption{Goodness of fit. We lists the number of degrees of freedom
(d.o.f.), the reduced $\chi^2_{eff}$, and the corresponding
probability of the best--fit model. $\Lambda$CDM models have 1342
degrees of freedom, uncoupled and vcDE models have 1341, while ccDE
models have 1340.}
\label{tab:chi2}
\vglue0.2truecm
\begin{center}
\begin{tabular}{cccc}
\hline \rule[-1ex]{0pt}{3.5ex} & $\chi^2_{eff}$ & prob.  \\ \hline dDE
& 1.062 & 5.2 \% \\ ccDE & 1.066 & 4.7 \% \\ vcDE & 1.074 & 2.9 \% \\
$\Lambda$CDM & 1.066 & 4.7 \% \\

\hline
\end{tabular}
\end{center}
\end{table}

\begin{figure}
\plotone{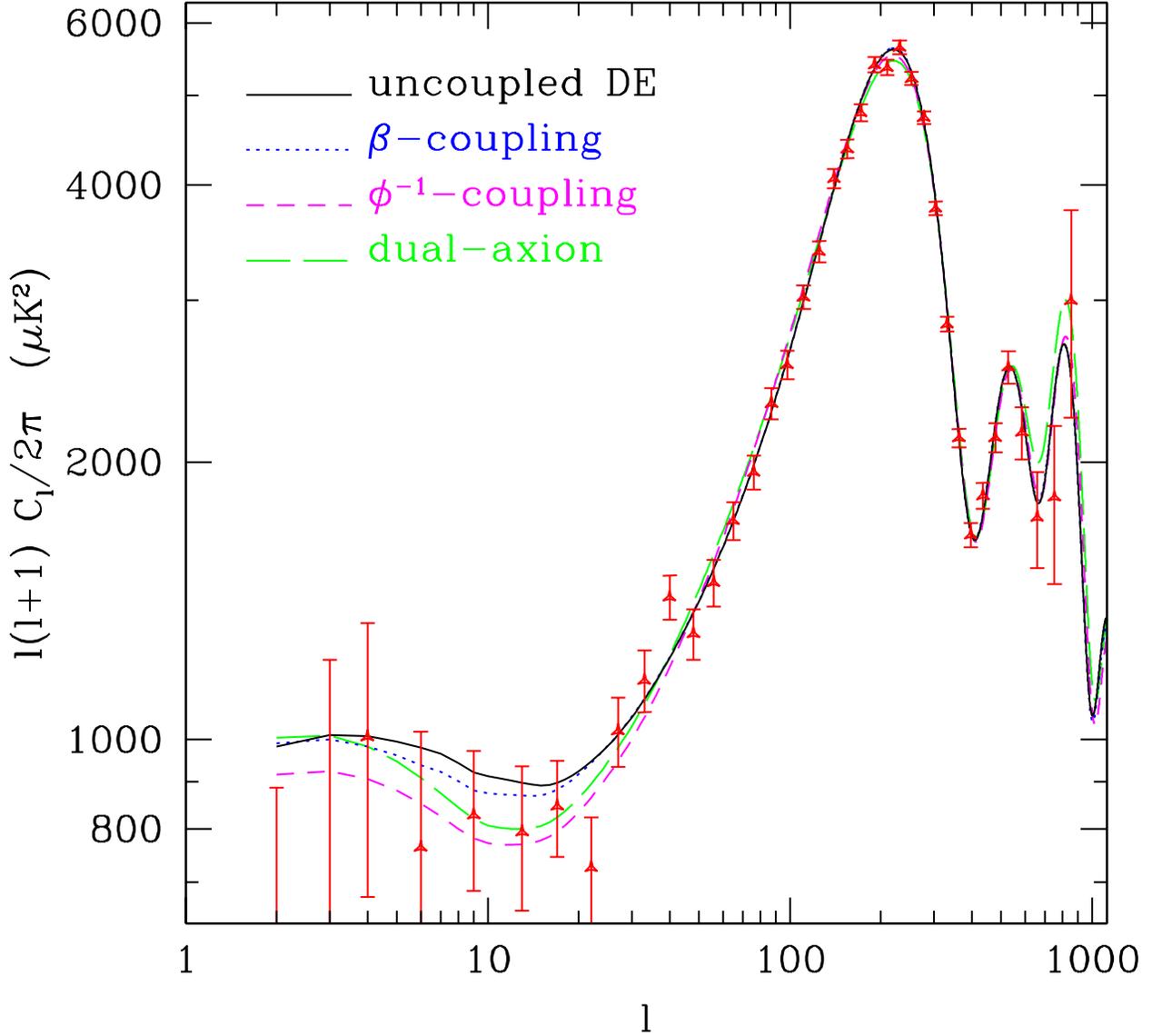}
\epsscale{1.0}
\caption{$C_l^T$ spectra for the best fit SUGRA (solid line), ccDE
(dotted line) and vcDE (dashed). Dual--axion (dot--dashed) models are
also shown; they can be considered a particular case within vcDE
models, for which $10 \le \lambda \le 10.5$ (see text). The binned
first--year WMAP data are also plotted.}
\label{fig:taps}
\end{figure}
\begin{figure}
\plotone{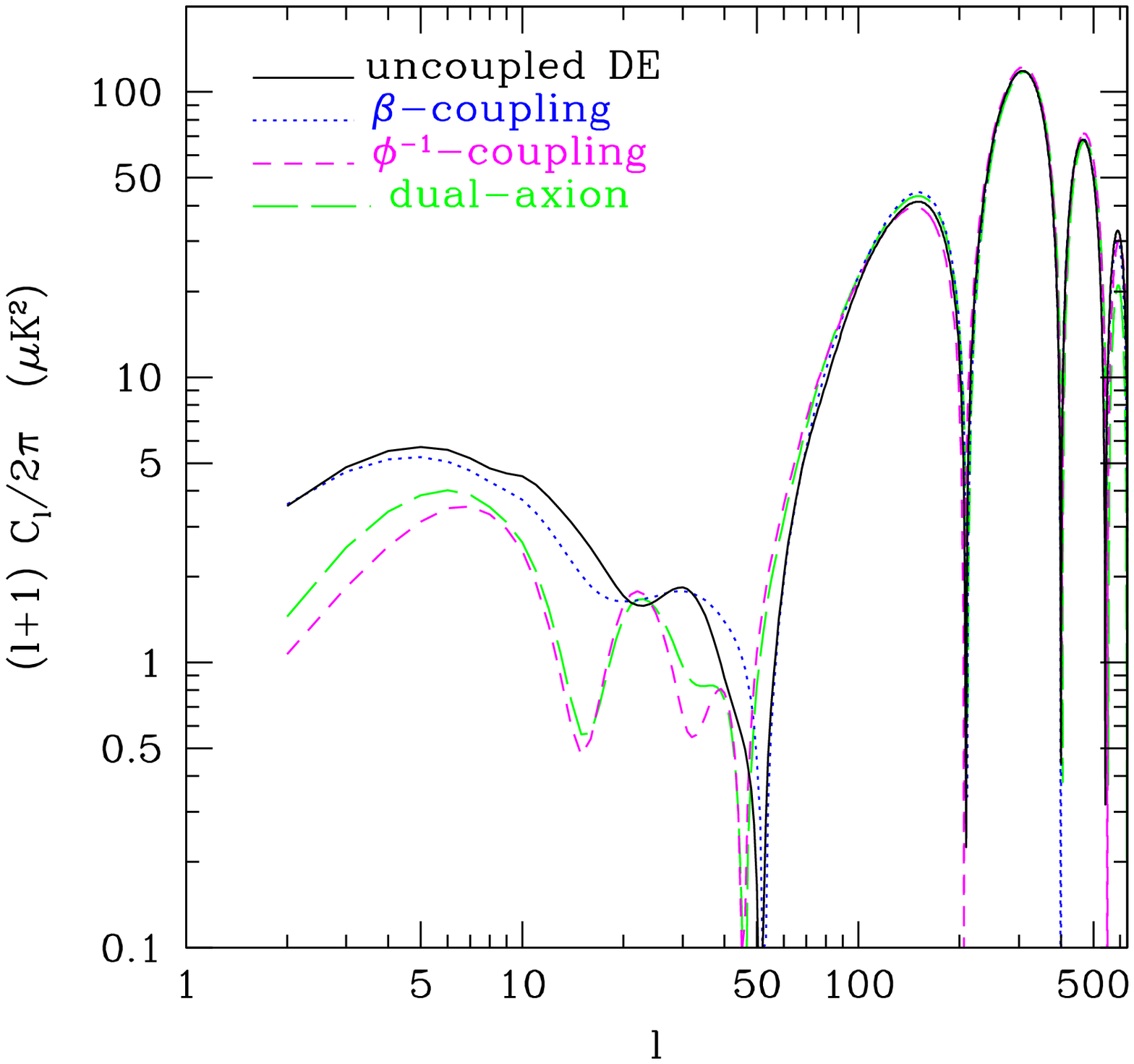}
\epsscale{1.0}
\caption{Best fit $C_l^{TE}$ spectra.}
\label{fig:teaps}
\end{figure}
\begin{figure}
\plotone{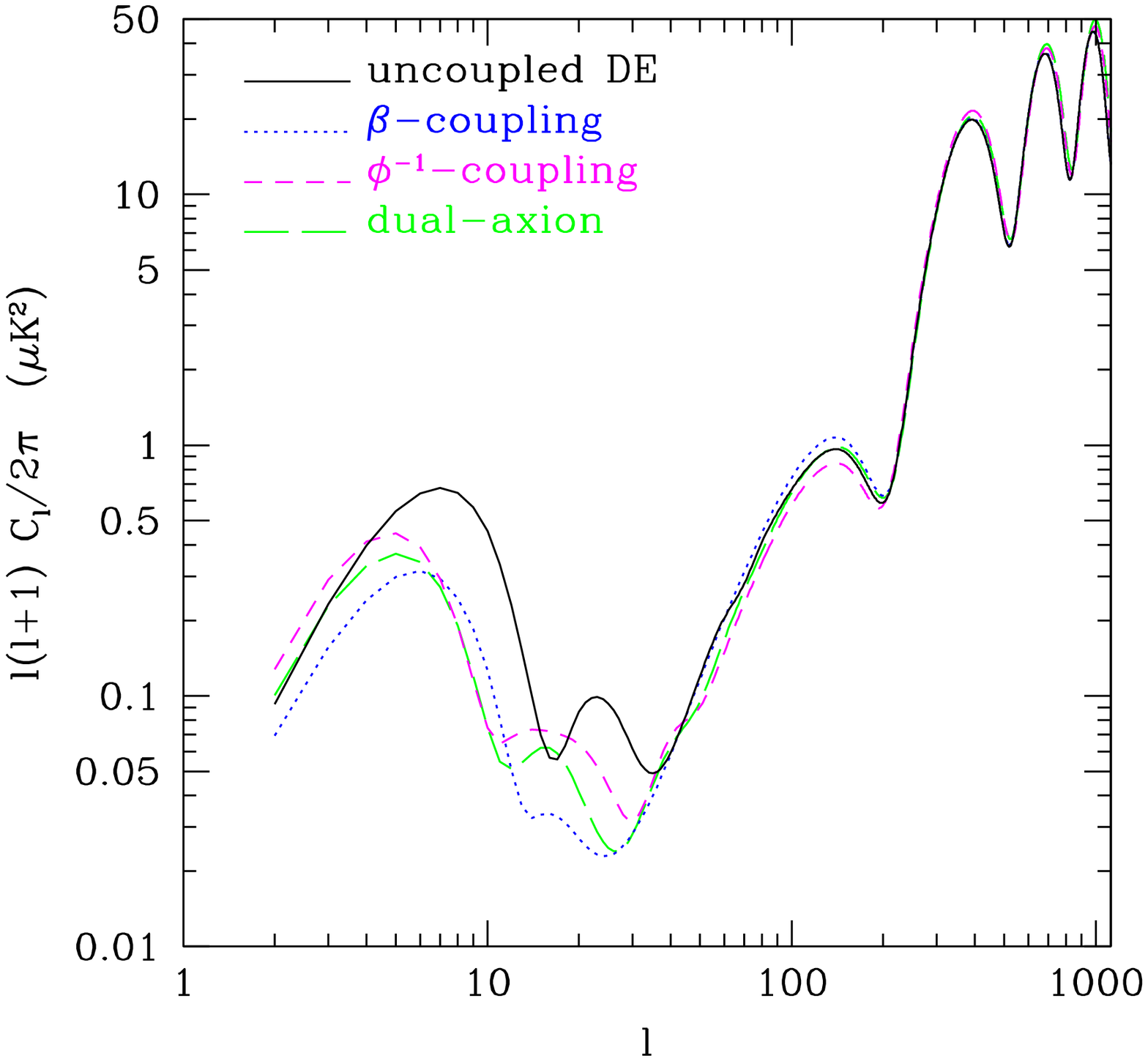}
\epsscale{1.0}
\caption{Best fit $C_l^{E}$ spectra.}
\label{fig:eaps}
\end{figure}

These constraints apparently arise from polarization data: when DE
begins to affect cosmic expansion, the $\phi$ evolution acts on DE
density and pressure, as well as on its coupling, and requires an
$l$--dependence of multipole amplitudes ablt to meet data only for a
limited range of $\Lambda$ values. This can be noticed in Figures
\ref{fig:taps} and \ref{fig:teaps}, where the best--fit $C_l^T$ and
$C_l^{TE}$ spectra for all best--fit models (apart of $\Lambda$CDM)
are compared.

Fig.~\ref{fig:taps} also shows why no model neatly prevails. At large
$l$ all best--fit models yield similar behaviors.  Discrimination
could be improved only through better large angular scale observations,
especially for polarization, able to reduce errors on small--$l$
harmonics.

\section{Discussion}
\subsection{Uncoupled SUGRA models}
SUGRA {\it uncoupled} models are found to be consistent with WMAP
data. The ratio $w = p/\rho$, for these models is typically $ \lesssim
- 0.80$ at $z=0$. However, they exhibit a fast variation of $w$, which
is already $\sim -0.6$ at $z \sim 1$--2. Such a decrease, however,
does not cause a conflict with data and {\it these models perform even
better than $\Lambda$CDM}. 

We also considered the effect of adding some priors. For uncoupled or
constant--coupling SUGRA models, the analysis in the presence of
priors leads to analogous conclusions. There are however variations in
the estimate of cosmological parameters. First, the opacity $\tau$ is
pushed to values exceeding the $\Lambda$CDM estimates (see also Corasaniti at al. 2004). This can be understood in two complementary ways:
(i) ISW effect is stronger for dDE than for $\Lambda$CDM, as the field
$\phi$ evolves during the expansion; hence, DE effects extend to a
greater redshift, increasing $C_l^T$ values in the low--$l$ plateau
\cite[e.g.,][]{weller}. To compensate this effect the fit tends to
shift $n_s$ greater values. Owing to the $\tau$--$n$ degeneracy, this
is then balanced by increasing $\tau$. (ii) It must also be reminded
that the expected TE correlation for dDE, at low $l$, is smaller
(Colombo et al.~2003). Then, when TE correlation data is included, a
given correlation level is fit by a greater $\tau$. In any case,
however, $\tau \sim 0.07$ keeps consistent with data within less than
2--$\sigma$'s.

Greater $\tau$'s affect also $\Omega_b h^2$ estimates, whose best--fit
value is then greater, although consistent with $\Lambda$CDM within
1--$\sigma$. Adding a prior on $\Omega_b h^2 = 0.0214 \pm 0.0020$
(BBNS estimates \citep{bbns}) lowers $h$, slightly below HST findings,
but still well within 1--$\sigma$. We therefore consider the effect
of adding a prior also on $h$. 

The effects of these priors are shown in Figures~\ref{fig:pdf1D1}
and~\ref{fig:pdf1D2}. The new distributions are given by the dashed
red line (prior on $\Omega_b h^2$) and the dot--dashed blue line
(prior on $h$). The former prior affects mainly $\tau$ and $n_s$;
$\tau$ and $n_s$ are lowered to match WMAP's findings, the high tail
of $\tau$ distribution is partly suppressed. The physical analysis of
primeval objects causing reionization
(Ciardi et al. 2003, Ricotti et al. 2004).
which are still allowed but certainly not {\it required}.

\subsection{ccDE SUGRA models}
The latter prior favors greater $h$ values. In the absence of
coupling, this favors low--$\lambda$ models, closer to $\Lambda$CDM.
In fact, the sound horizon at decoupling is not affected by the energy
scale $\Lambda$, while the comoving distance to last scattering band
is smaller for greater $\lambda$'s. Then, for greater $\lambda$'s
lower $h$ values are favored, so to meet the first peak. In the
presence of coupling, there is a simultaneous effect on $\beta$:
greater $\beta$'s yield a smaller sound horizon at recombination, so
that the distribution on $h$ is smoother.

A previous analysis of WMAP limits on constant coupling models had
been carried on by Amendola \& Quercellini (2003). Their analysis
concerned potentials $V$ fulfilling the relation $dV/d\phi = BV^N$,
with suitable $B$ and $N$. Furthermore, they assume that $\tau \equiv
0.17$.  Our analysis deals with a different potential and allows more
general parameter variations. The constraints on $\beta$ we find are
less severe. It must be however outlined that $\beta \simgreat
0.1$--0.2 seem however forbidden by a non--linear analysis of
structure formation (Macci\`o et al. 2004).

\subsection{vcDE SUGRA models}
The main peculiarity of $\phi^{-1}$--models is that, although the
$\xi^2$ value is not much worsened in respect to other cases,
parameters are more strongly constrained in this caseq. This is
evident in Fig.~\ref{fig:pdf2D3}. In particular, at variance from the
former case, the energy scale $\Lambda$ is significantly constrained.

Several other parameters are constrained, similarly to coupled or
uncoupled SUGRA models. What is peculiar of $\phi^{-1}$--models
is the range of $h$ values which turn out to be favored. The 
best--fit 2--$\sigma$ interval does not extend much below 0.85$\, $.

This is a problem for these models in their present form. Should new
data support a greater $h$ value, vcDE models would however be
favored. Quite in general, however, we must remind that vcDE models
were considered here only in association with a SUGRA potential; vcDE
models with different potentials can possibly agree with a smaller
$h$.

\subsection{Dual axion models}
A particular but significant case, among vcDE models, are dual axion
models (DAM; for a detailed analysis see Mainini \& Bonometto 2004,
Colombo et al 2004).

These models arise from a generalization of the PQ solution of the
strong--$CP$ problem (Peccei \& Quinn 1977), obtained by replacing the
NG potential acting on a complex scalar field $\Phi$ with a
quintessence potential admitting a tracker solution. While, in the
usual PQ approach and its generalization, only the phase $\theta$ of
the $\Phi$ field is physically sgnificant, in the DAM also its modulus
$\phi$ is physically observable. In both cases, quanta of the $\theta$
field are axions and, if DM is made of axions, its density parameter
$\Omega_{o,m}$ is set either by the $F_{PQ}$ parameter, in the NG
potential, or by the energy scale $\Lambda$, in the quintessence
potential. In the latter case, therefore, both DM and DE features
follow from fixing the single parameter $\Lambda$.

A detailed analysis of DAM shows that it belongs to the set of vcDE
cosmologies; its main peculiarity is that $\lambda$ and $\Omega_{o,m}$
are no longer independent degrees of freedom.  Henceforth, this model
is strongly constrained and depends on the same number of parameters
as SCDM. By varying $\Omega_{o,m}$ in its physical range, $\lambda$
keeps however close to 10. This great value of $\lambda$ therefore
causes a conflict with the observed $h$ range. All previous comments
on vcDE models however hold, while it is also possible that extra
contribution to DM, arising from topological singularities expected in
axion theories, can ease the $h$ problem.

\section{Conclusions}
The first evidences of DM are fairly old; they date some 70 years ago.
However, only in the late Seventies limits on CMB anisotropies showed
that non--baryonic DM had to be dominant. DE can also be dated back to
Einstein's {\it cosmological constant}, although only SN Ia data
revived it, soon followed by data on CMB and deep galaxy samples.

The main topic of this paper is the fit of WMAP data with various DE
models: $\Lambda$CDM, SUGRA dynamical and constant coupling DE models,
as well as variable coupling DE models. In the last model DM and DE
are coupled in a non--parametric way, with $C = \phi^{-1}$.

The fits of WMAP data to these models yield similar $\chi^2$'s. At
variance from other model categories, however, in vcDE models, CMB
data constrain $\Lambda$. This is due to the stronger effects of
$\phi$ variations on the detailed ISW effect, as they affect both DE
pressure and energy density, as well as DE--DM coupling.  In
principle, this strong impact of $\phi$ variation could badly disrupt
the fit and make vcDE models significantly farther from data. It is
noticeable that this does not occur.

The success of vcDE models would be complete if the favored range of
values of the Hubble parameter ($h \sim 0.8$--1) could be slightly
lowered. This range is however obtained just for a SUGRA potential.
Furthermore, primeval fluctuations were assumed to be strictly
adiabatic while, in axion models, a contribution from isocurvature
modes can be expected. This could legitimately affect the apparent
position of the first peak in the anisotropy spectrum, so completing
the success of the model, in a fully self--consistent way.

\begin{acknowledgements}
\noindent
{\bf Acknowledgements}.
This work was partially supported by the PRIN project {\it Astroparticles}
of the Italian MIUR.

\end{acknowledgements}

\end{document}